\documentclass[12pt]{iopart}

\expandafter\let\csname equation*\endcsname\relax
\expandafter\let\csname endequation*\endcsname\relax

\usepackage{graphicx}
\usepackage{amsmath}
\usepackage{amssymb}
\usepackage{microtype}

\begin{document}

\title[Self-ordering and cavity cooling using a train of ultrashort pulses]{Self-ordering and cavity cooling using a train of ultrashort pulses}

\author{Valentin Torggler$^1$, Ivor Kre\v si\' c$^{2,3}$, Ticijana Ban$^2$, Helmut Ritsch$^1$}

\address{$^1$Institut f\"ur Theoretische Physik, Universit\"at Innsbruck, Technikerstra{\ss}e 21a, A-6020 Innsbruck, Austria}
\address{$^2$Institut za fiziku, Bijeni\v cka cesta 46, HR-10000 Zagreb, Croatia}
\address{$^3$Institut f\"ur Theoretische Physik, TU Wien, Wiedner Hauptstra{\ss}e 8-10/136, A-1040 Vienna, Austria}

\ead{helmut.ritsch@uibk.ac.at}
\vspace{10pt}
\begin{indented}
\item[]December 2019
\end{indented}

\begin{abstract}
A thin atomic gas in an optical resonator exhibits a phase transition from a homogeneous density to crystalline order when laser illuminated orthogonal to the resonator axis. We study this well-known self-organization phenomenon for a generalized pumping scheme using a femtosecond pulse train with a frequency spectrum spanning a large bandwidth covering many cavity modes. We show that due to simultaneous scattering into adjacent longitudinal cavity modes the induced light forces and the atomic dynamics becomes nearly translation-invariant along the cavity axis. In addition the laser bandwidth introduces a new correlation length scale within which clustering of the atoms is energetically favorable. Numerical simulations allow us to determine the self-consistent ordering threshold power as function of bandwidth and atomic cloud size. We find strong evidence for a change from a second order to a first order self-ordering phase transition with growing laser bandwidth when the size of the atomic cloud gets bigger than the clustering length. An analysis of the cavity output reveals a corresponding transition from a single to a double pulse traveling within the cavity. This doubles the output pulse repetition rate and creates a new time crystal structure in the cavity output. Simulations also show that multi-mode operation significantly improves cavity cooling generating lower kinetic temperatures at a much faster cooling rate.
\end{abstract}

%
%
%
%
%

\section{Introduction}
\label{sec:introduction}

Laser light forces on a dilute atomic gas are enhanced and qualitatively modified within optical resonators as the back-action of the atoms onto the field dynamics cannot be ignored \cite{domokos2003mechanical}. In particular, even at large detuning, the dipole force does not simply induce a conservative optical potential, but the dynamic nonlinear atom-field coupling opens a wealth of new phenomena \cite{Ritsch2013cold}. An important consequence is the possibility for opto-mechanical cooling without spontaneous emission via the delayed response of the cavity field to the atomic motion which can be applied to any kind of polarizable particles \cite{horak1997cavity,domokos2002collective,schleier2011optomechanical,wolke2012cavity,hosseini2017cavity,delic2019cavity}.

As another surprising phenomenon one finds phase transitions towards regular order and improved collective cooling if the atoms are transversely illuminated by a sufficiently strong pump laser tuned just below the eigenfrequency of a cavity mode \cite{domokos2002collective,chan2003observation}. This spontaneous ordering process can be traced back to collective (superradiant) coherent Bragg scattering from an emerging periodic atomic density distribution forming self-consistently at the interference maxima of pump and cavity field \cite{domokos2002collective}. Such spontaneous crystallization can still be observed at zero temperature as a quantum phase transition from a superfluid to a supersolid order \cite{nagy2010dicke,baumann2010dicke,Ritsch2013cold,mekhov2012quantum}.

When several laser frequencies and field modes are simultaneously applied, the corresponding atom-field dynamics gets much more complex. Recently ultracold atomic gases in cavities have been used to simulate or emulate a wide class of other exotic solid state phenomena like edge states, topological insulators or quasi-crystal formation \cite{mivehvar2017superradiant,mivehvar2019emergent} or general Hamiltonians for quantum enhanced simulated annealing \cite{torggler2019quantum,torggler2017quantum}. Using atoms with internal spin, cavity mediated interactions have the potential to implement a large class of lattice spin models \cite{landini2018formation,Mivehvar2019cavity,kroeze2019dynamical,kroeze2018spinor}.

In several theoretical proposals it was predicted that cavity cooling is strongly enhanced when the number of modes involved is enlarged \cite{domokos2002dissipative,nimmrichter2010master}. Interestingly the use of only two distinct cavity modes and laser frequencies already strongly enlarges the complexity of theoretical modeling. While a single cavity mode creates a virtually infinite range all-to-all interaction throughout the whole cavity volume, by help of a large number of longitudinal modes with corresponding driving frequencies, one should be able to engineer interaction forces of a very general spatial shape \cite{torggler2017quantum}. While this sounds extremely complex in practice, one can make use of the fact that cavity modes like comb lines of ultra short pulse trains are equidistant. Hence frequency matching of only two lines will automatically align a huge manifold of frequency components. In general the frequency distribution of the pump light then gives a direct handle on the spatial shape of the interaction forces \cite{holzmann2016tailored}.

Here we generalize the pump light to frequency combs (FCs) with a large bandwidth as generated by a pulse train coming from a mode-locked femtosecond laser. The locking of two comb lines to two different longitudinal modes is then sufficient to guarantee resonant overlap of all modes over virtually the whole pump bandwidth. This way the atoms can simultaneously coherently scatter thousands of different frequencies into the cavity, which can form a newly shaped pulse train between the mirrors. The average optical potential over this pulse train then determines the spatial order of the atoms in a self-consistent way. As in the case of a single-mode the atoms mostly tend to order in such a way that they maximize the total intracavity intensity. This can lead to local periodic order together with long-range clustering of the atoms. Interestingly, the temporal shape of the output pulses and their repetition rate provides for nondestructive observation of the atomic ordering process. As the output pulse train dynamically acquires a different shape and a higher repetition rate than the input pulse, the ordering phenomenon should be closely related to a time crystal formation.

This paper is organized as follows: We first introduce the proposed set-up and a semi-classical mathematical model, which includes a discussion about basic physical properties in Sec.\ \ref{sec:model}. Next, in Sec.\ \ref{sec:stationary} we identify the lowest energy stationary states of the system and numerically calculate the ordering threshold using a self consistent mean-field approach. The dynamics for a wide range of operating parameters and initial conditions are discussed in Sec.\ \ref{sec:dynamics}. A special emphasis is put on the form of the ordered patterns and their signature in the cavity output light. Finally we discuss the extra benefits for cavity cooling in such a wide bandwidth multi-mode system. We conclude our work with an outlook on future perspectives in Sec.\ \ref{sec:conclusion}.

\section{Model}
\label{sec:model}

\begin{figure}
\centering
\includegraphics[width=0.5\columnwidth]{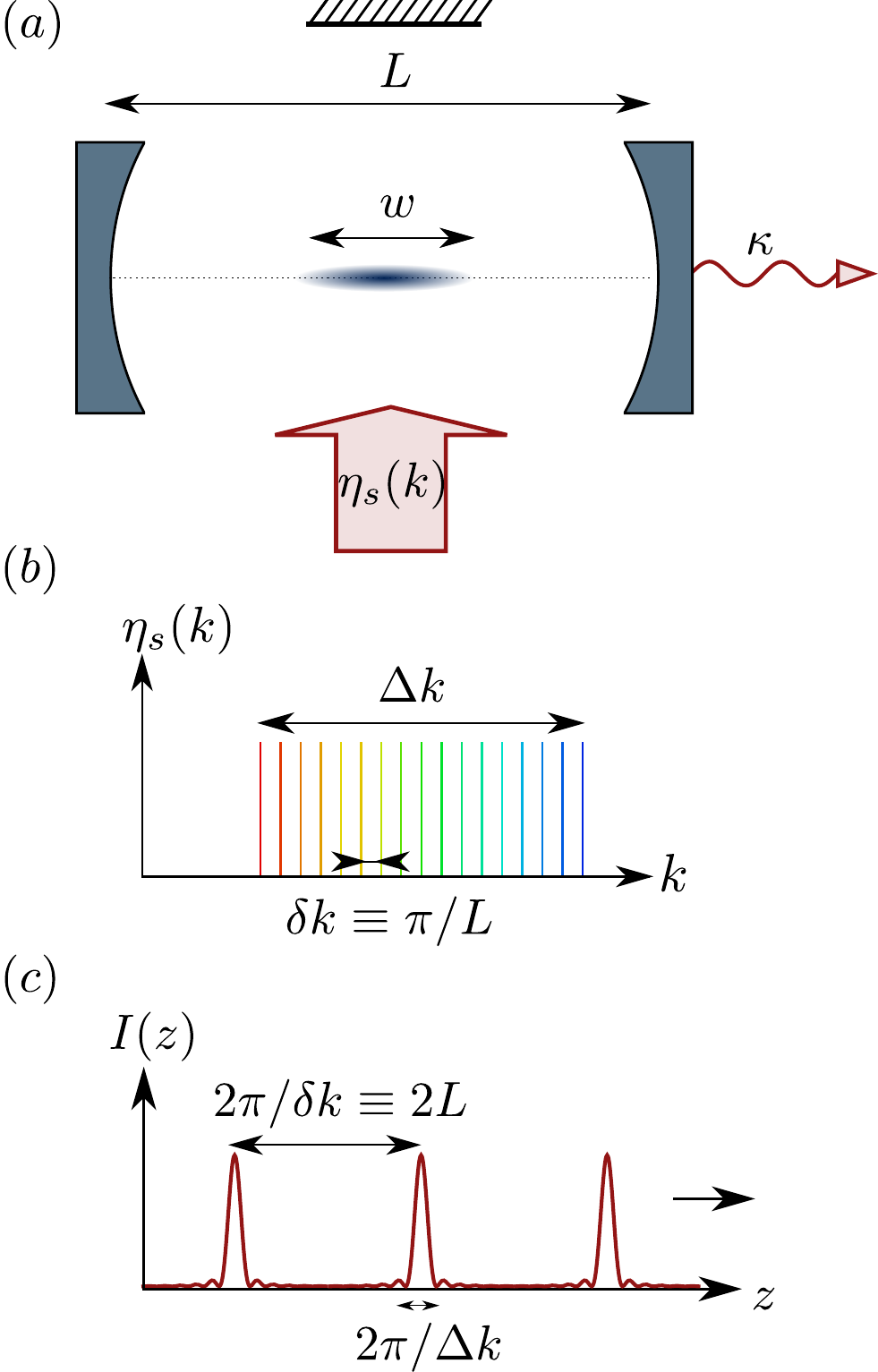}
\caption{(a) Sketch of the atomic cloud of size $w$ at the central axis of an optical resonator of length $L$, which is transversely pumped by a frequency comb. (b) The line spectrum $\eta(k)$  of the laser beam is specified by its bandwidth $\Delta k$ and line spacing $\delta k$. In order to match to the cavity modes we choose $\delta k \equiv \pi / L$. (c) This line spectrum corresponds to a train of phase-stabilized ultrashort pulses in the time domain, which translates to a regular train of pulses with width $2\pi/\Delta k$ and distance $2\pi / \delta k \equiv 2L$ in space, along the FC propagation axis.}
\label{fig:model}
\end{figure}

Let us consider an atomic cloud of $N$ atoms of mass $m_a$ trapped along the central axis within a high finesse Fabry-P\'erot cavity of length $L$ supporting a wide range of non-degenerate longitudinal modes. The modes have wave-vectors  $k_{m} = m \delta k$ and frequency distance $\delta k  = \pi / L$, where the mode indices $m$ range from $m_\mathrm{low}$ to $m_\mathrm{high}$. The atoms are transversally illuminated with a retro-reflected FC creating a standing wave perpendicular to the cavity as depicted in Fig.\ \ref{fig:model}(a). The laser frequency spectrum consists of equidistant narrow lines with spatial frequencies $k_{p,m}$ with a bandwidth $\Delta k$ [see Fig.\ \ref{fig:model}(b)]. The line spacing is chosen as $\delta k$, such that the round-trip time of a pulse inside the cavity and the pulse repetition period of the ultrashort pulse train match [see Fig.\ \ref{fig:model}(c)]. Pairs of cavity and FC modes can be tuned to be quasi-resonant up to a constant detuning $\Delta_c = \omega_{p,m} - \omega_{m}$ with the cavity frequencies $\omega_m = c k_m$ and the speed of light $c$. This detuning is adjusted to a value smaller than zero $\Delta_c < 0$. Due to the quasi-resonance, light is scattered from the FC into the cavity via the atoms, where the coupling between pump and cavity mode pairs $\eta$ is assumed to be constant in $m$. It depends on the pump strength, the atom-cavity coupling and the atomic detuning \cite{domokos2002collective}. Finally, the so-created cavity field leaks out of the resonator with the rate $\kappa$, which we assume to be independent of $m$ for simplicity.

The atoms are confined in a narrow dipole trap along the cavity axis ($x$ axis), where all pump mode functions have a maximum. The theoretical description is thus reduced to a 1D model. Moreover we assume that the atomic cloud is distributed around the center of the cavity at $x=0$ and has a size $w$, which is much smaller than the cavity length $L$. Then the cavity mode functions can be approximated by alternating cosines and sines, which we write as $\mathcal{G}_{c,m}(x) =\cos(k_m x - \phi_m)$ with $\phi_m = 0$ for odd $m$ and $\phi_m = \pi/2$ for even $m$.

\subsection{Mathematical model and its properties}
When the temperature of the atomic cloud is larger than the recoil energy, i.e.\ $k_\mathrm{B} T \gg \hbar \omega_\mathrm{R}^{k_m} = (\hbar k_m)^2/(2 m_a)$ for all $m$, the coupled dynamics of the dilute gas and the field is well described by the semi-classical stochastic differential equations \cite{domokos2002collective}
\begin{subequations}\label{eq:semiclassical}
\begin{align}
\dot x_j &= \frac{p_j}{m_a} \\
\dot p_j &= - \hbar \sum_m \eta [\partial_{x_j} N \Theta_m] (\alpha_m + \alpha_m^*) \label{eq:semiclassical_momentum} \\ 
\dot \alpha_m &= (i \Delta_c - \kappa) \alpha_m - i \eta N \Theta_m + \sqrt{\kappa} \xi_m \label{eq:semiclassical_fields}
\end{align}
\end{subequations}
coupling atomic positions $x_j$, momenta $p_j$ and cavity fields $\alpha_m$. The diffusion due to photon loss is taken into account by the Wiener process $\xi_m$, which obeys the noise correlations $\langle \xi_m(t) \xi_n^*(t') \rangle = \delta(t-t') \delta_{mn}$. The crucial quantities describing the amount of scattering into the $m$-th cavity mode are the atomic order parameters
\begin{equation}\label{eq:op_frequcomb}
\Theta_m = \frac{1}{N} \sum_{i=1}^N \cos(k_m x_i - \phi_m).
\end{equation}
We neglect here the dynamical Stark shift of the cavity field and assume that the mode spacing is large compared to the coupling strength and cavity detuning such that there is no cross-scattering between different cavity modes \cite{keller2018quenches}. 

In the single-mode case, it is well known that the described set-up and Eqs.\ \eqref{eq:semiclassical} lead to a stationary state for $\Delta_c < 0$ \cite{niedenzu2011kinetic}. Below a threshold of the laser power, the stationary spatial distribution of the atoms is homogeneous, while above this threshold it exhibits a density modulation with the period of one wavelength \cite{domokos2002collective}. With the rescaled pump intensity \cite{niedenzu2011kinetic,schuetz2015thermodynamics}
\begin{equation}\label{eq:zeta}
\zeta = \frac{4 N \eta^2 \Delta_{c}^2}{(\Delta_{c}^2+\kappa^2)^2},
\end{equation}
the threshold is at $\zeta = 1$ for an atomic cloud at the self-consistent stationary temperature
\begin{equation}\label{eq:temperature}
k_\mathrm{B} T_\mathrm{st} = \hbar \frac{\Delta_c^2+\kappa^2}{-4 \Delta_c}.
\end{equation}
The stationary momentum distribution is well approximated by a thermal state with the self-consistent stationary temperature $T_\mathrm{st}$ as long as $|\Delta_c|/\omega_\mathrm{R} \gg 1$ and $|\Delta_c|/\omega_\mathrm{R} \gg 4 \zeta$ \cite{niedenzu2011kinetic}, i.e.\ when the pump intensity is not too high. This sometimes allows for employing equilibrium thermodynamics to describe the stationary properties of the system, even though the system is open \cite{schuetz2015thermodynamics}. When the cavity parameters are tuned such that the stationary temperature is smaller than the initial temperature, the dynamics cools the atomic cloud -- a phenomenon called cavity cooling, which allows for sub-Doppler cooling without relying on the internal structure of the atoms \cite{horak1997cavity}.

Within the thermal limit, the stationary momentum distribution is not altered when pumping two modes as long as $\kappa$ and $\Delta_c$ are uniform \cite{keller2017phases}. In such a case the definition of Eq.\ \eqref{eq:temperature} provides a unique stationary temperature. The thresholds for self-ordering and the spatial form of the stationary states, however, strongly depend on the mode structure \cite{torggler2014adaptive}. Moreover, the type of the phase transition can change from second to first order, resulting in quasi-stationary states depending on the initial spatial condition and temperature \cite{keller2018quenches}.

In this work we present a further generalization to the two-mode case by pumping many adjacent modes, which have a similar frequency. Since there is no cross-scattering between the modes, the extension from the two-mode case is theoretically straightforward, the sums simply run over larger range. Thus the results about the stationary temperature can be generalized to many modes for uniform $\kappa$ and $\Delta_c$. The mode structure, however, is obviously changed leading to an alteration of the thresholds and the spatial stationary distribution.

\subsection{Frequency comb and mode regimes}
We now aim to identify the crucial properties and parameters of the mode structure of a frequency comb pumped cavity. The spectrum of the frequency comb can be formally written as a convolution of an envelope function $\eta_\mathrm{env}(k)$ and a Dirac comb. In units of the pump-cavity coupling it takes the general form
\begin{equation}\label{eq:frequencycomb}
\eta_s(k) = \eta_\mathrm{env}(k) \ast \sum_{n=-\infty}^\infty \delta(k - n\delta k)
\end{equation}
with $(f \ast g)(k) = \int_{-\infty}^\infty f(k') g(k-k') \mathrm{d} k'$. For our choice of uniform coupling $\eta$, the envelope function is a rectangle with height $\eta$ and width $\Delta k$ centered around $k_c$. The spatial shape of the laser pulses is related to the envelope via a Fourier transform and thus has a width of about $2\pi / \Delta k$, a pulse-to-pulse distance $2\pi / \delta k$ and a characteristic sinc form [see Fig.\ \ref{fig:model}(c)], where $\mathrm{sinc}(x) = \sin(x) / x$.

\begin{figure}
\centering
\includegraphics[width=0.6\columnwidth]{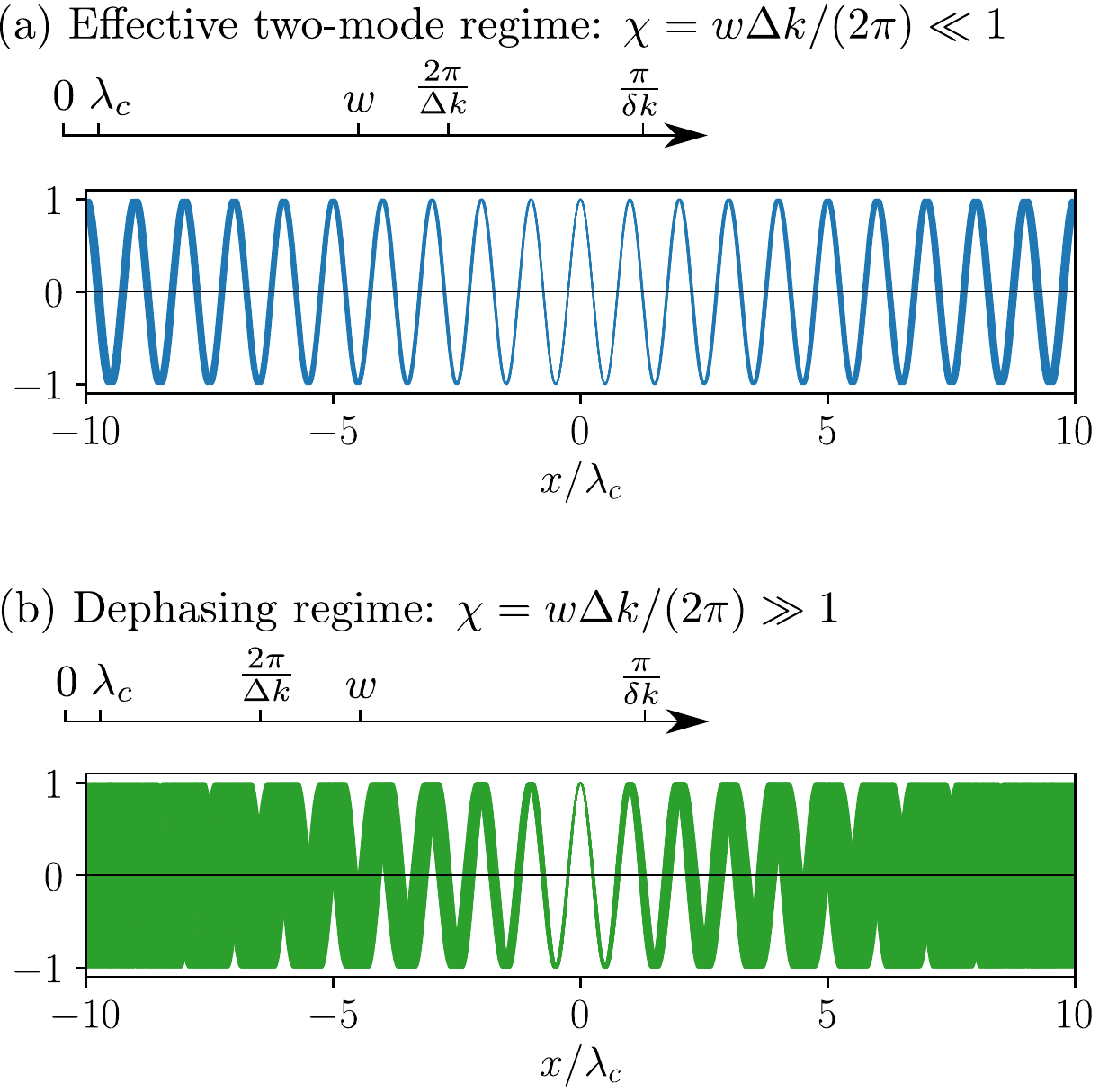}
\caption{Qualitative sketch of the different FC bandwidth regimes. The smallest length scale is given by the comb line wavelengths  as e.g. the central comb wavelength $\lambda_c=2\pi/k_c$. The beat length between neighboring comb lines is $\pi / \delta k \equiv L$, which typically is much larger than the atomic cloud of size $w$. In the effective two-mode regime (a)  $2\pi / \Delta k$ is larger than the atomic cloud $w$ and thus none of the involved modes differs significantly within the atomic cloud. The blue curve shows the fields of $8$ adjacent cosine modes over the interval $w = 20 \lambda_c$ with $2\pi / \Delta k \approx 45.9 \lambda_c$. (b) Dephasing regime: Distant modes run out of phase within the atomic cloud size and the spatial laser pulse width is smaller than the atomic cloud. The green curves show $50$ cosine modes with $2\pi / \Delta k \approx 9.2 \lambda_c$, which clearly dephase within interval $w = 20 \lambda_c$. The mode spacing $\delta k$ is the same in both cases. For visual clarity we depict only the cosine modes, omitting sine modes.}
\label{fig:mode_regimes}
\end{figure}

In this work the spectral lines of the FC match all longitudinal cavity modes within a certain range, leading to the mode spacing $\delta k = \pi/L$ and a pulse distance of one cavity round trip length $2L$. Put differentily, this means that the (spatial) shapes of \textit{adjacent} modes dephase only on a length scale of the cavity length $L$, and since $w \ll L$ they do not differ within the atomic cloud. Thus adjacent sine-cosine mode pairs can be assumed to effectively have the same frequency. Within this ``dense'' mode parameter regime $w \delta k / \pi \ll 1$, the behavior is left to depend on the pulse width in relation to the atomic cloud size, i.e.\ on the parameter
\begin{equation}
\chi = \frac{w \Delta k}{2\pi}.
\end{equation}
For $\chi \ll 1$, the spatial pulse width is large compared to the cloud. In this regime \textit{all} modes have the same shape within the extent of the atomic cloud, and thus can be effectively assumed to have the same frequency [see Fig.\ \ref{fig:mode_regimes}(a)]. In the center of the cavity there are thus two groups of spatially coinciding modes, $M/2$ sine modes and $M/2$ cosine modes. Since the physics can be effectively described by two modes, we refer to this regime as the \textit{effective two-mode regime}.

For $\chi \gtrsim 1$ the spatial pulse width becomes smaller than the atomic cloud size and thus introduces a new length scale to the system. Modes which are far apart in the spectrum spatially dephase within the atomic cloud [see Fig.\ \ref{fig:mode_regimes}(b)], and thus we refer to this regime as the \textit{dephasing regime}.

In the following we will investigate these two regimes and the transition from one to the other by tuning $\chi$. Practically, this is done by increasing the bandwidth while keeping the mode density and the atomic trap size constant, or by increasing the size of the atomic cloud keeping the bandwidth constant.

\section{Stationary states}
\label{sec:stationary}
In this section we consider the form of the stationary states and aim to find the threshold for different $\chi$. For this we first introduce the stationary potentials. The stationary cavity fields
\begin{equation}\label{eq:fields}
\alpha_m = \frac{\eta}{\Delta_c + i \kappa} N \Theta_m
\end{equation}
are attained after a time $\kappa^{-1}$ and can be calculated from Eq.\ \eqref{eq:semiclassical_fields} by setting $\dot \alpha_m = 0$ and omitting the noise. Replugging this expression into Eq.\ \eqref{eq:semiclassical_momentum} yields an adiabatic force, which can be integrated to the single-particle potential
\begin{equation}\label{eq:single-particle_pot}
U_\mathrm{MF}(x;\Theta_1,...,\Theta_M) = -2 k_\mathrm{B} T_\mathrm{st} \zeta \sum_m \Theta_m(x_1,...,x_N) \cos(k_m x - \phi_m).
\end{equation}
This is a mean-field potential for a single particle at position $x$ created by all particles via $\Theta_m(x_1,...,x_N)$. Moreover, summing over all particles reveals the total potential energy \cite{keller2018quenches}
\begin{equation}\label{eq:potentialenergy}
U(x_1,...,x_N) = -k_\mathrm{B} T_\mathrm{st} N \zeta \sum_m \Theta_m^2 = -k_\mathrm{B} T_\mathrm{st} N M \zeta {\bar \Theta}^2,
\end{equation}
where the square of the total order parameter
\begin{equation}\label{eq:thetarms}
\begin{aligned}
{\bar \Theta}^2 = \frac{1}{M} \sum_{m=m_\mathrm{low}}^{m_\mathrm{high}} \Theta_m^2
\end{aligned}
\end{equation}
is the crucial quantity describing ordering to all modes. A minimization of the potential energy thus corresponds to a maximization of the square of the total order parameter.

\subsection{Low energy states far above threshold}
\label{ssec:stationarystates}

For a high pump power $\zeta$ far above the self-ordering threshold, the potential depth is large compared to the stationary temperature $T_\mathrm{st}$. The stationary spatial distribution then minimizes the potential energy $U$ and maximizes the square of the total order parameter ${\bar \Theta}^2$. It is thus worthwhile taking a closer look at this quantity.

As discussed in the previous section, adjacent cosine-sine pairs have the same effective frequency within the atomic cloud. In the expression in Eq.\ \eqref{eq:thetarms} we can thus summarize these adjacent pairs using the identity $\cos(\alpha)\cos(\beta) + \sin(\alpha)\sin(\beta) = \cos(\alpha - \beta)$ for each pair, and obtain
\begin{equation}
{\bar\Theta}^2 \approx \frac{1}{2MN^2} \sum_{i,j} \sum_{m=m_\mathrm{low}}^{m_\mathrm{high}} \cos(k_m (x_i-x_j)),
\end{equation}
where the factor $1/2$ emerges since we still sum over all $M$ modes. Rewriting this expression as a convolution of a rectangular envelope and a Dirac comb in (spatial) frequency space as in Eq.\ \eqref{eq:frequencycomb}, further yields the expression
\begin{equation}\label{eq:thetarms_sinc}
{\bar\Theta}^2 \approx \frac{1}{2N^2} \sum_{ij} \mathrm{sinc}\left(\pi \frac{x_i - x_j}{w} \chi\right) \cos(k_c (x_i - x_j)).
\end{equation}
Here we made an approximation by considering only a single sinc, which is very accurate since the next sinc is shifted by $2\pi/\delta k = 2L$ and is thus far outside of the atomic cloud. Note that in the dense mode regime the total order parameter is bound by $\bar \Theta \leq 1/\sqrt{2}$. This is a consequence of the fact that the atoms cannot simultaneously order to a sine and a cosine mode with the same wavelength.

The multi-mode behaviour enters via the sinc depending on the parameter $\chi$. In the effective two-mode regime $\chi \ll 1$, the sinc is always approximately $1$. A low energy distribution is thus left to maximize the cosine, which favors ordering of the atoms in a $\lambda_c$-periodic grating, as in the single-mode case. The absolute position in space, however, is not fixed since only the distances of particles enter. This translational invariance comes from the coupling to sine and cosine mode pairs with the same (approximate) wavelength. Considering one mode pair only, the translation over one wavelength of an atomic distribution would then shuffle photons from the sine mode to the cosine mode and back, while leaving the potential energy unchanged. This resembles the U(1) symmetry described for crossed cavities \cite{leonard2017supersolid} and atoms in ring cavities \cite{gangl2000cold}. Let us emphasize that in the present case this only valid close to the center of the cavity and when the dynamical Stark shift of the individual modes can be neglected.

In addition to the $\lambda_c$-periodic grating, a localization of the atomic cloud to some width smaller than the pulse width $2\pi / \Delta k$ becomes energetically favorable in the dephasing regime ($\chi > 1$). In such a way the atoms stay in the in-phase region of the modes [see Fig.\ \ref{fig:mode_regimes}(b)] and the sinc takes a value close to one. A wider distribution could scatter only a narrower bandwidth and thus less power.

\subsection{Self-ordering threshold from a mean-field model}
\label{ssec:thresholds}
We now aim to find the threshold of the total laser pump power $\zeta_\mathrm{tot} = M\zeta$ for self-ordering at the self-consistent temperature $T_\mathrm{st}$ as a function of $\chi$. Remembering that the stationary state can be approximated by a thermal state in typical parameter regimes \cite{schuetz2015thermodynamics}, we assume that the stationary state is a Boltzmann distribution with the temperature $T_\mathrm{st}$ and the mean-field potential given in Eq.\ \eqref{eq:single-particle_pot}. To obtain the self-consistent cavity fields creating this potential, we employ the method introduced in Ref.\ \cite{asboth2005self} for the single-mode case. After choosing an initial spatial atomic distribution $P_0(x)$, one calculates its order parameters $\Theta_m[P_\mathrm{0}]$, which can be written as
\begin{equation}
\Theta_m[P] = \int_{-L}^{L} \mathrm dx \, P(x) \cos(k_m x-\phi_m)
\end{equation}
analogously to Eq.\ \eqref{eq:op_frequcomb}. These specify a mean-field potential $U_\mathrm{MF}(x; \Theta_1[P_0],...,\Theta_M[P_0])$ created by the distribution $P_0$. The updated atomic configuration $P_1$ in the next iteration is the normalized Boltzmann distribution in this potential with the temperature $T_\mathrm{st}$ [Eq.\ \eqref{eq:temperature}]. In this vein the $n$-th step of the algorithm can be formulated as
\begin{equation}\label{eq:iteration}
P_{n+1}(x) \propto \exp\left(-\frac{U_\mathrm{MF}(x; \Theta_1[P_n],...,\Theta_M[P_n])}{k_\mathrm{B} T_\mathrm{st}}\right).
\end{equation}
The iteration is then repeated for $N_\mathrm{iter}$ times, until the distribution does not change anymore.

Slow ramping of the laser power over the threshold can be emulated by the following procedure. For a fixed $\chi$ we start with a low pump strength below the threshold and a homogeneous distribution, and calculate the stationary state using the above described algorithm [Eq.\ \eqref{eq:iteration}]. In the next step, we use this stationary state as initial state for the iteration with a slightly higher pump strength. Continuing this way we simulate an up-sweep of the laser power. Analogously, for the down-sweep we start with the stationary state obtained from the up-sweep far above the threshold and then reduce the pump power in small steps.

\begin{figure}
\centering
\includegraphics[width=0.6\columnwidth]{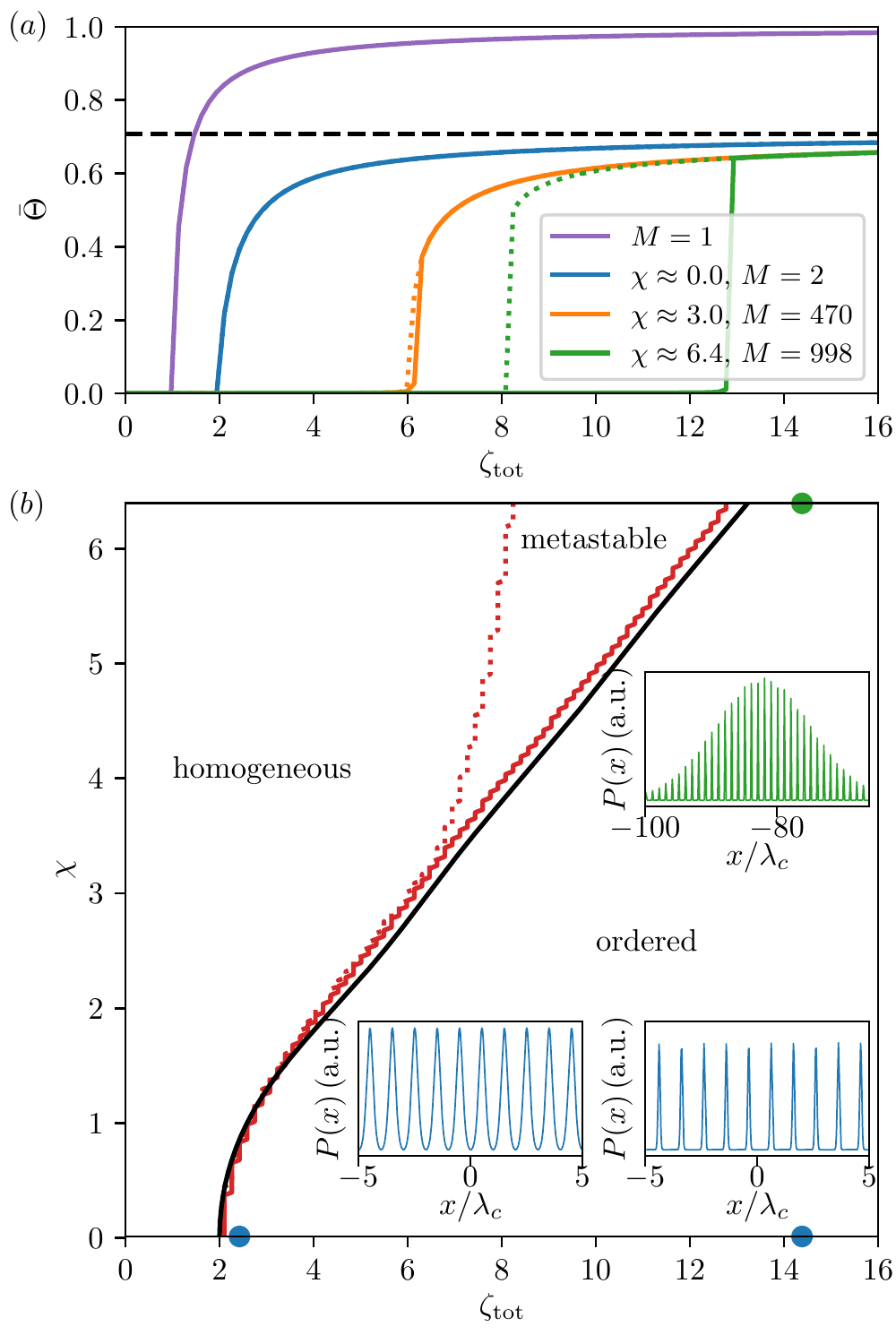}
\caption{Stationary states obtained from the self-consistent iteration [Eq.\ \eqref{eq:iteration}]. (a) The total order parameter $\bar \Theta$ of the stationary state as a function of the pump intensities $\zeta_\mathrm{tot} = M \zeta$. The solid (dashed) lines depict the results for an upward-sweep (downward-sweep) of the laser intensity. In the two-mode case ($\chi \approx 0$), these curves coincide and we get the a typical second order self-organization phase transition. In the multi-mode regime for larger $\chi$, jumps in the order parameter and a hysteresis as signs of a first order phase transition appear. (b) The corresponding phase diagram as a function of $\chi$ and the laser power $\zeta_\mathrm{tot}$. The red curves depict the thresholds for an upward-sweep (solid line) and a downward-sweep (dotted line) of the laser power. The black curve is a theoretical prediction of the phase boundary depending on the mode structure given in Eq.\ \eqref{eq:threshold}. In the large $\chi$-regime there is a metastable region and hysteresis. The insets depict atomic distributions at different positions of the phase diagram above the threshold. Apart from the wavelength grating, the distribution is flat for small $\chi$ (blue) and is localized for large $\chi$ (green).}
\label{fig:phase_diagram}
\end{figure}

The atomic distribution is confined to a finite interval with the width $w = 300 \lambda_c$ centered around the cavity center. The parameter $\chi$ is varied by changing the bandwidth $\Delta k$ while keeping the mode spacing (or mode density) $\delta k$ constant. The maximal bandwidth we consider is $\Delta k_\mathrm{max} / k_c \approx 0.0213 = 2.13\%$ with $M=998$. Correspondingly, the mode spacing is given by $\delta k / k_c = \Delta k_\mathrm{max}/k_c/998 = 2.13 \cdot 10^{-5}$, which is deep in the dense mode regime ($w \delta k / \pi \approx 0.013 \ll 1$). Note that in theory the maximum number of modes (here $998$) can be chosen quite arbitrarily as long as the dense mode condition $w \delta k / \pi \ll 1$ is fulfilled, while in the experiment it is fixed by the comb-cavity set-up. The cavity parameters are $\kappa = 400 \omega_\mathrm{R}^{k_c}$ and $\Delta_c = - \kappa$ and the stationary temperature hence has the simple form $k_\mathrm{B} T_\mathrm{st} = \hbar \kappa / 2$.

As shown in the insets of Fig.\ \ref{fig:phase_diagram}(b), above the threshold the atomic distributions have a form similar to the low energy states maximizing Eq.\ \eqref{eq:thetarms} discussed in the previous section \ref{ssec:stationarystates}. In the effective two-mode regime ($\chi \ll 1$), the $\lambda_c$-periodic density modulation extents over the whole interval and the peaks get narrower for larger pump intensities (blue). In the dephasing regime ($\chi > 1$), the atoms additionally localize within an interval smaller than the pulse width, which is $46.9 \lambda_c$ for $\chi \approx 6.4$ (green).

Interestingly, for small $\chi$ the order parameter continuously grows at the transition point, and the results from up- and down-sweep coincide [see Fig.\ \ref{fig:phase_diagram}(a)]. For larger $\chi$, in contrast, the order parameter exhibits a jump at the transition point and the threshold depends on the sweep direction. This hysteresis hints that the type of the phase transition changes from second order to first order upon increasing $\chi$.

The threshold $\zeta_\mathrm{tot}^c$ increases as a function of $\chi$ (see Fig.\ \ref{fig:phase_diagram}), signifying that more total input power is needed to create a sufficiently deep potential to stabilize an ordered pattern for larger $\chi$. The reason for this is that upon increasing the bandwidth, less modes are being scattered into by an atomic distribution with a fixed width and less power is transmitted to the cavity. In other words, at some $\chi$ not the whole spectrum can be transmitted anymore.

In this vein, in order to find an expression for the multi-mode threshold, we rescale the single-mode threshold $\zeta^c = 1$ by a factor representing the efficiency of power transmission into the cavity compared to the single-mode case. This efficiency factor is given by the square of the total order parameter ${\bar \Theta}^2[P_c]$, where $P_c$ is an atomic distribution which optimally couples to a single-mode, i.e.\ maximizes the cosine in Eq.\ \eqref{eq:thetarms_sinc}. That is, $P_c$ is a $\lambda_c$-periodic grating with some arbitrary width. In this quite general sense the multi-mode threshold is modeled by
\begin{equation}
\zeta_\mathrm{tot}^c = {\bar\Theta}^{-2}[P_c].
\end{equation}
A rather simple expression can be derived for the up-sweep. Coming from below the threshold, the atoms are distributed over the whole interval, and thus $P_c$ has the width $w$. Explicitly, we write
\begin{equation}
P_c(x) = \frac{1}{N_\lambda} \sum_{j = -N_\lambda/2+1}^{N_\lambda/2} \delta(x-j \lambda_c)
\end{equation}
with $N_\lambda = w / \lambda_c$. With this distribution, the estimated multi-mode threshold becomes
\begin{equation}\label{eq:threshold}
\zeta_\mathrm{tot}^c = \frac{2 N_\lambda^2}{\sum_{i,j = -N_\lambda/2+1}^{N_\lambda/2} \mathrm{sinc}(\pi   \chi (i-j) /N_\lambda)}.
\end{equation}
This quantity is depicted as a black line in Fig.\ \ref{fig:phase_diagram}(b) and very well resembles the results from the self-consistent iteration.

The situation for the down-sweep is different, since by coming from above the threshold the width of $P_c$ depends on $\chi$. Figure \ref{fig:phase_diagram} shows the numerical results for the down-sweep boundary. Again, for larger $\chi$, the up- and down-sweep phase boundaries depart from each other, spanning a region where several metastable states exist.

\section{Dynamics}
\label{sec:dynamics}

In addition to the analytical and stationary considerations above, we now simulate the dynamics of the frequency comb set-up by numerical integration of the stochastic differential equations of motion given in Eqs.\ \eqref{eq:semiclassical}.

\subsection{Self-ordering}
\label{ssec:self-ordering}

First we compare the dynamics of $N=100$ atoms for different pump strengths for the initial temperature $k_\mathrm{B} T = \hbar \kappa / 2$. The initial spatial distribution is uniform on an interval of the width $w_\mathrm{ini}$, which we choose as either $20\lambda_c$ or $300\lambda_c$. Note that the situation in the dynamical case here is different to the self-consistent calculations above since the particles can move out of the interval and due to the relatively small particle number.

Again we choose the cavity parameters as $\kappa = 400 \omega_\mathrm{R}^{k_c}$ and $\Delta_{c} = -\kappa$. The bandwidth $\Delta k_\mathrm{max} / k_c \approx 2.13\%$ is spanned by $50$ modes now, leading to the mode spacing $\delta k / k_c = \Delta k_\mathrm{max}/k_c/50 = 4.26 \cdot 10^{-4}$. That means that we are still in the high mode density regime where adjacent modes do not dephase significantly.

\begin{figure}
\centering
\includegraphics[width=0.6\columnwidth]{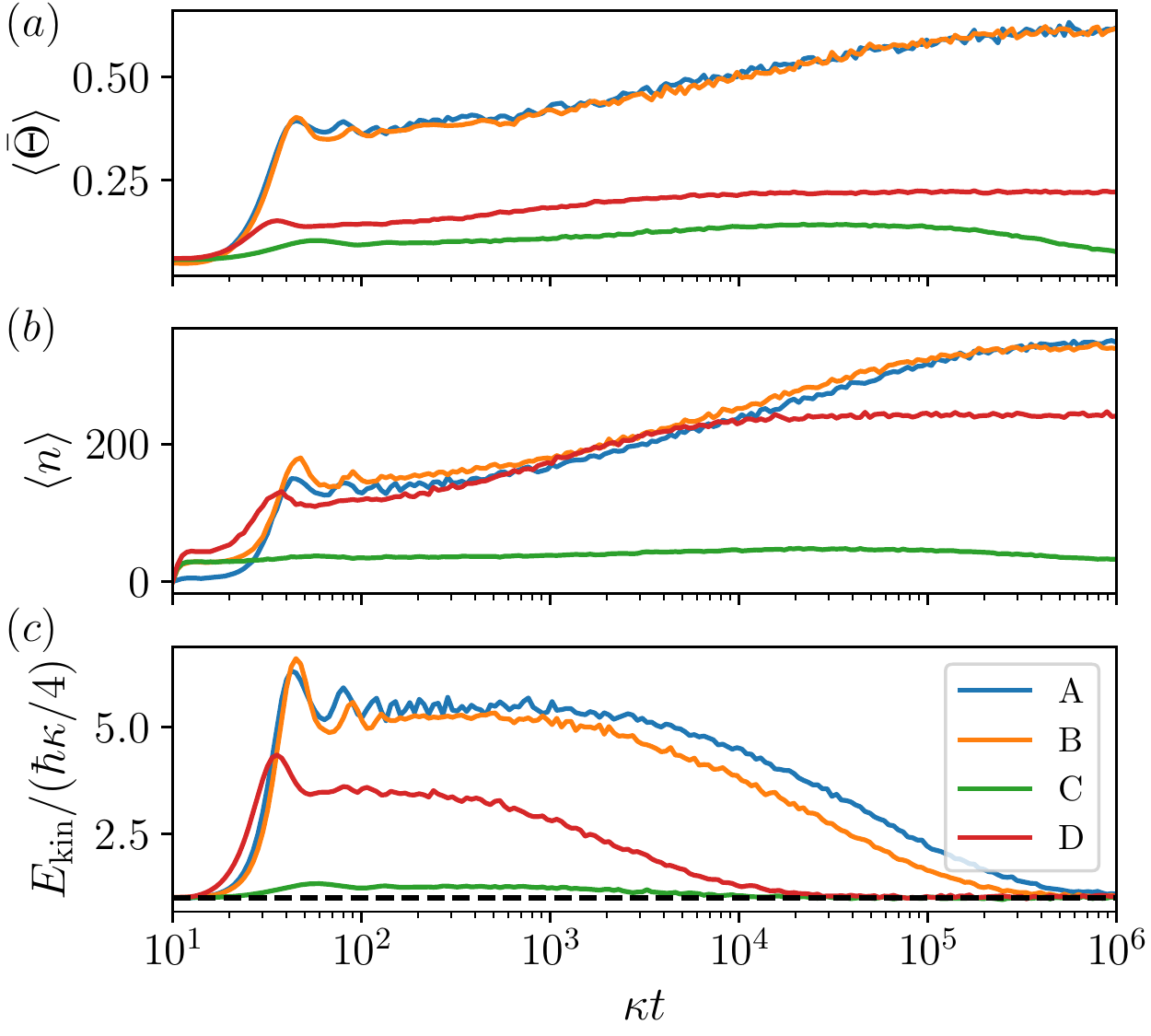}
\caption{Dynamic evolution of (a) the total order parameter $\langle \bar \Theta \rangle$ for N=100 particles averaged over $50$ trajectories with (b) the corresponding total intra-cavity photon number $\langle n \rangle$ and (c) the average kinetic energy $E_\mathrm{kin} = \langle p^2 \rangle / (2m_a)$ in units of the predicted stationary value $\hbar \kappa / 4$. The dashed line is set at 1. The blue and and orange curve are similar and correspond to the total pump intensity $\zeta_\mathrm{tot} = 15$ and the initial width of the atomic cloud $w_\mathrm{ini} = 20 \lambda_c$ for $M=2$ (A, blue) and $M=50$ (B, orange). Increasing the initial width to $w_\mathrm{ini} = 300 \lambda_c$ with $M=50$ and $\zeta_\mathrm{tot} = 15$ leads to the green curve (C). The red curve (D) corresponds to pumping above the single-mode threshold, $\zeta_\mathrm{tot} = 60$ ($\zeta = 1.2$). Note the logarithmic time axis.}
\label{fig:dynamics_avg}
\end{figure}

\begin{figure}
\centering
\includegraphics[width=0.6\columnwidth]{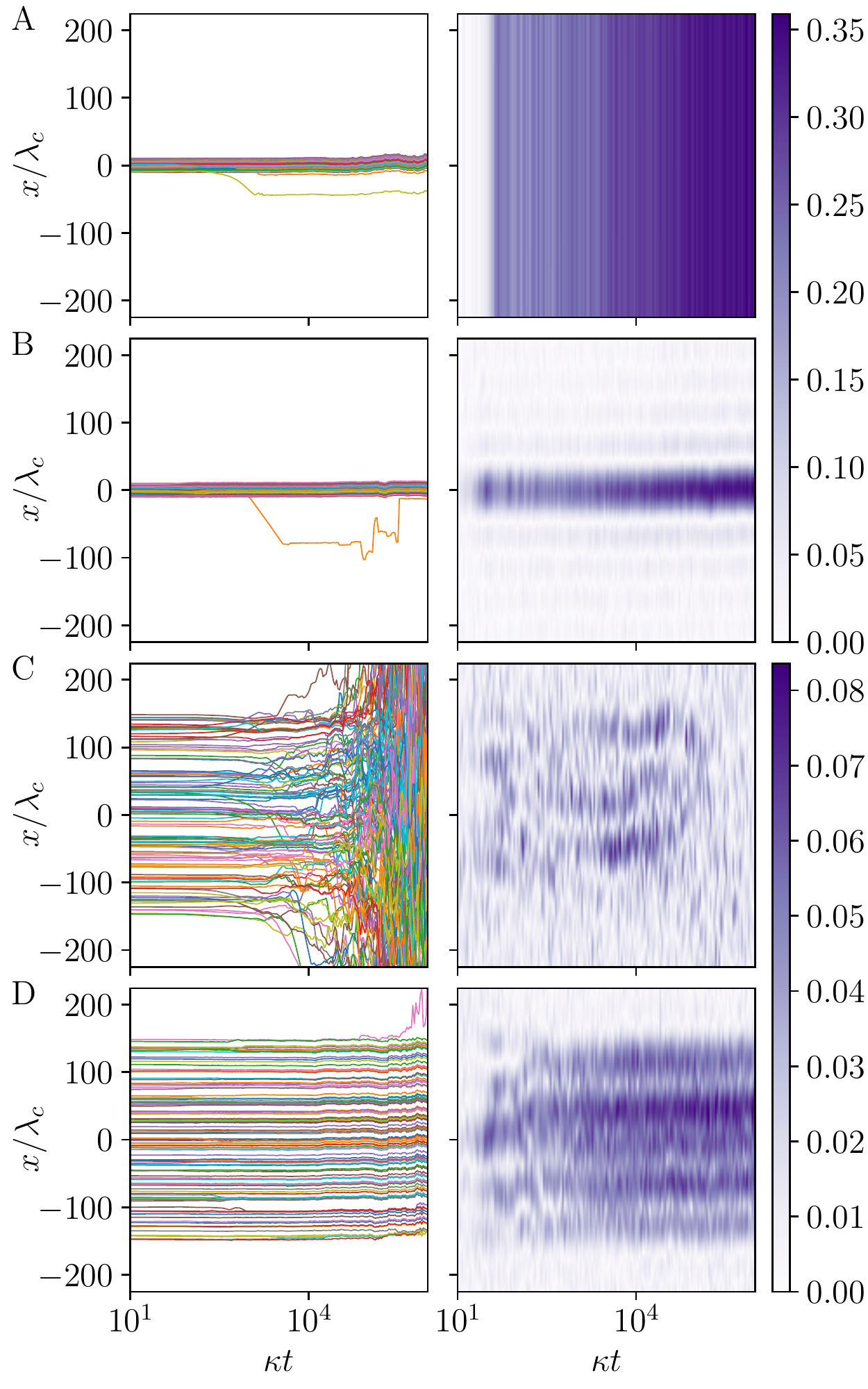}
\caption{Dynamics of the spatial atomic distribution for example trajectories for the parameters in Fig.\ \ref{fig:dynamics_avg} denoted by (A), (B), (C) and (D). The left column depicts the atomic positions $x$ as a function of time. The right column shows the smoothed field distribution (normalized to 1) in the cavity $\mathcal{F}(x)$ [Eq.\ \eqref{eq:F_env}], which gives information about where the atoms are ordered in space for a sufficiently large bandwidth. Note that the field intensity distribution is uniform in the two-mode case $M=2$.}
\label{fig:dynamics_pos}
\end{figure}

\begin{figure}
\centering
\includegraphics[width=0.6\columnwidth]{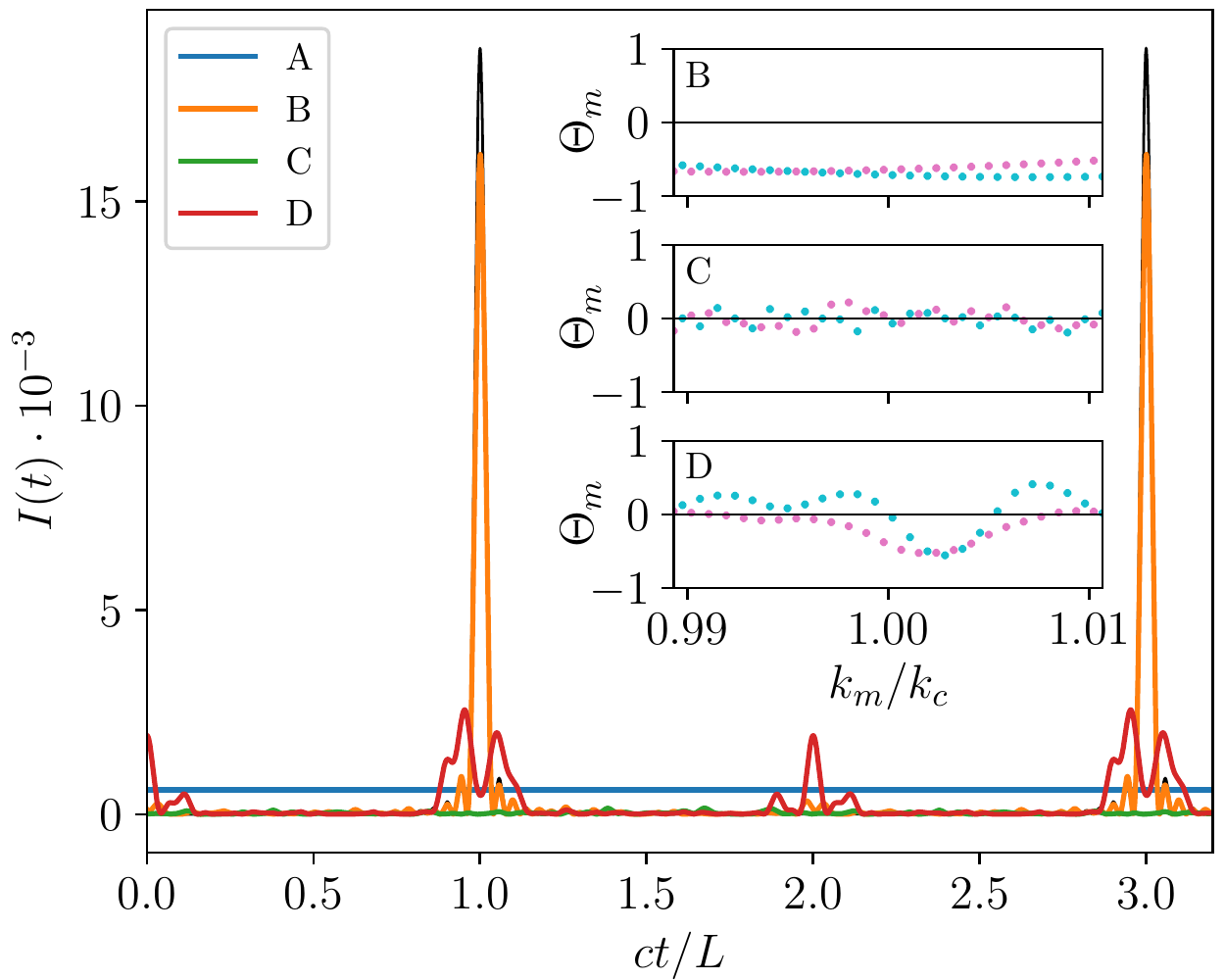}
\caption{Short time dependence of the cavity output intensity at $\kappa t = 10^6$ in Fig.\ \ref{fig:dynamics_pos} in units of the time light needs to travel one cavity length $L$. (A) A continuous wave output is observed for $M=2$ as also the input is continuous ($\Delta k \approx \delta k$). In the other cases (B-C), the input is pulsed ($M=50$, $\Delta k \gg \delta k$). (B) In the case of an atomic distribution localized in the cavity center, these pulses are transmitted to the cavity without changing their shape significantly. The black curve shows the input pulse shape for reference. (C) The weakly ordered case leads to a weak cavity output with irregular aperiodic structure. (D) A self ordered atomic distribution results in periodic pulses of different shape representing the intracavity order. Since ordering happens at several locations the output pulses have multiple peaks breaking the pump pulse symmetry. Insets show the order parameters for a range of cosine (magenta) and sine (cyan) modes.}
\label{fig:pulses}
\end{figure}

Figure \ref{fig:dynamics_avg} depicts the trajectory averages of the total order parameter $\langle \bar \Theta \rangle$, the total intra-cavity photon number $\langle n \rangle  = \sum_m \langle |\alpha_m|^2 \rangle$ and the kinetic energy $E_\mathrm{kin} = \langle p^2 \rangle / (2 m_a)$. For relatively free particles it is related to the temperature by $E_\mathrm{kin} = k_\mathrm{B} T / 2$ due to the equipartition theorem. In Fig.\ \ref{fig:dynamics_pos} the position of the particles is plotted for an example trajectory. Moreover, we depict the averaged field distribution inside the cavity
\begin{equation}\label{eq:F}
F(x) = \frac{1}{M} \sum_m \Theta_m \cos(k_m x - \phi_m),
\end{equation}
with the normalized spatial form of the $m$-th cavity mode field $\Theta_m \cos(k_m x - \phi_m)$. The envelope of this quantity gives insight about where in space the particles are ordered and can be extracted as the smoothed field distribution
\begin{equation}\label{eq:F_env}
\mathcal{F}(x) = \sqrt{F^2(x) \ast \mathrm{rect}\left(\frac{x}{\lambda_c}\right)},
\end{equation}
with the rectangle function $\mathrm{rect}(x) = 1$ for $-1/2 < x < 1/2$ and $0$ else. From our simulations we obtain the time evolution of the trajectory averaged order parameters $\langle \Theta_m \rangle(t)$, which enables us to consider the time evolution of the function $\mathcal{F}(x)$. Note that also experimentally the order parameters can be indirectly measured via a measurement of the cavity output fields. When the dynamics of the cavity fields $\alpha_m$ is much faster than the atomic motion, we can use the stationary approximation of the fields from Eq.\ \eqref{eq:fields} to obtain an approximate expression for the order parameters
\begin{equation}\label{eq:op_quadr}
\Theta_m \approx \frac{\sqrt{\Delta_c^2+\kappa^2}}{N|\eta|} X_\theta
\end{equation}
in terms of the measurable field quadratures
\begin{equation}
X_\theta = \mathrm{Re}\left[\alpha_m \exp\left(-i\mathrm{arg}[(\Delta_c+i\kappa)^{-1}]\right)\right].
\end{equation}
Finally, Fig.\ \ref{fig:pulses} depicts the laser output intensity
\begin{equation}
I(t) = \left|\sum_m \alpha_m e^{-i \omega_m t}\right|^2
\end{equation}
for the trajectories in Fig.\ \ref{fig:dynamics_pos} at the final time $\kappa t = 10^6$.

We start with the case of two modes for comparison, labeled (A) in Figs.\ \ref{fig:dynamics_avg}, \ref{fig:dynamics_pos} and \ref{fig:pulses}. The total and single-mode pump strength are $\zeta_\mathrm{tot} = 15$ and $\zeta = 7.5$, respectively, where the latter is much larger than one and thus far above threshold. The dynamics can be split into two time regions: The fast initial increase of the order parameter $\langle\bar \Theta\rangle$ and the photon number $\langle n\rangle$ is accompanied with an increase of the kinetic energy. Here the atoms fall into their self-created potential. On a much longer time scale, the atomic cloud is cooled down to the approximate stationary temperature (dashed line), leading to further localization in the potential and thus a further increase of the order parameter. This characteristic dynamics resembles the single mode case and has been discussed in detail e.g.\ in Ref.\ \cite{schuetz2014prethermalization}. In contrast to the single-mode case, the potential is nearly translationally invariant, which allows for a collective translational motion of the ordered atoms [see Fig.\ \ref{fig:dynamics_pos}(A)]. The field distribution $\mathcal{F}(x)$ is spatially uniform for two modes (cosine-sine pair), since the light fields do not contain any information about the absolute position. Just as the cavity input light, the output light is continuous for two modes and two comb lines (see Fig.\ \ref{fig:pulses}).

Case (B) shows the multi-mode scenario ($M=50$) for the same total pump intensity $\zeta_\mathrm{tot} = 15$ ($\zeta = 0.3$) and a small initial width $w_\mathrm{ini}=20\lambda_c$. Since $\chi_\mathrm{ini} = w_\mathrm{ini} \Delta k / (2\pi) \approx 0.426$ is well below one, the initial distribution is narrower than the pulse width and the atoms are in the in-phase region. Due to this and since the pump strength is above the multi-mode threshold, the atoms immediately order into a (approximately) $\lambda_c$-periodic grating, which hinders the atomic cloud from expanding. The atoms thus stay in the in-phase region and the curves of the collective quantities in Fig.\ \ref{fig:dynamics_avg} closely resemble the two-mode case. When an atom leaves the in-phase region, it is not trapped anymore, in contrast to the two-mode case. Also the field distribution $\mathcal{F}(x)$ is not uniform anymore, but is localized on the dephasing length scale. The cavity output field phases thus include information about the absolute position of the ordered atoms. As depicted in Fig.\ \ref{fig:pulses}, the cavity output is pulsed and the shape of the input pulse is not considerably changed by the Bragg reflection on the atoms.

The situation drastically changes for an initial distribution of $w_\mathrm{ini} = 300 \lambda_c$, which is much wider than the pulse width ($\chi_\mathrm{ini} = w_\mathrm{ini} \Delta k / (2\pi) = 6.39 > 1$). Case (C) depicts this situation with the same parameters as in (B). Initially, there is only a small increase in the order parameter since the atomic cloud is too wide to order to all modes. This comes hand in hand with a smaller increase of the kinetic energy, and faster cooling, which will be examined in the next section. The potential depth is too small to permanently trap the atoms, and does not hinder the expansion of the atomic cloud (see Fig.\ \ref{fig:dynamics_pos}). With increasing (dynamical) width of the cloud, less and less modes can couple to the cloud and the total order parameter starts to decrease again. At intermediate times the field distribution shows several regions of ordering, which later disappear.

Finally in (D) we consider a very large pump intensity $\zeta_\mathrm{tot} = 60$ above the single-mode threshold ($\zeta = 1.2$). In spite of the wide initial distribution $w_\mathrm{ini} = 300 \lambda_c$, the cloud organizes into some stable pattern, since even a single-mode can sustain ordering here. This initial freezing does not only impede the expansion, but also further localization of the atoms. As depicted in Fig.\ \ref{fig:dynamics_pos} and the inset in Fig.\ \ref{fig:pulses}, the atomic cloud orders to different modes in different points in space. This leads to modified output pulse shapes, as depicted in Fig.\ \ref{fig:pulses}.

Summarizing, we find that the initial width of the distribution is the main factor deciding whether strong ordering to many modes occurs. Initial widths smaller than the dephasing length often result in ordering which can be stable on large time scales even below the multi-mode threshold obtained using the self-consistent iteration. For initial widths larger than the dephasing length instead, the atoms do not find the multi-mode stationary state in the cases we considered and order only to few modes simultaneously. Even for pump strengths slightly above the multi-mode threshold obtained before, the trapping is not strong enough to stabilize the atoms. Finally, above the single-mode threshold, where one mode is sufficient to sustain ordering, the atoms freeze close to their current position. A better ordering also in the case of wide initial distributions might be obtained by slowly ramping up the pump strength to its final value or by using higher initial temperatures \cite{keller2018quenches}.

\subsection{Cavity cooling}
\label{ssec:cooling}

\begin{figure}
\centering
\includegraphics[width=0.6\columnwidth]{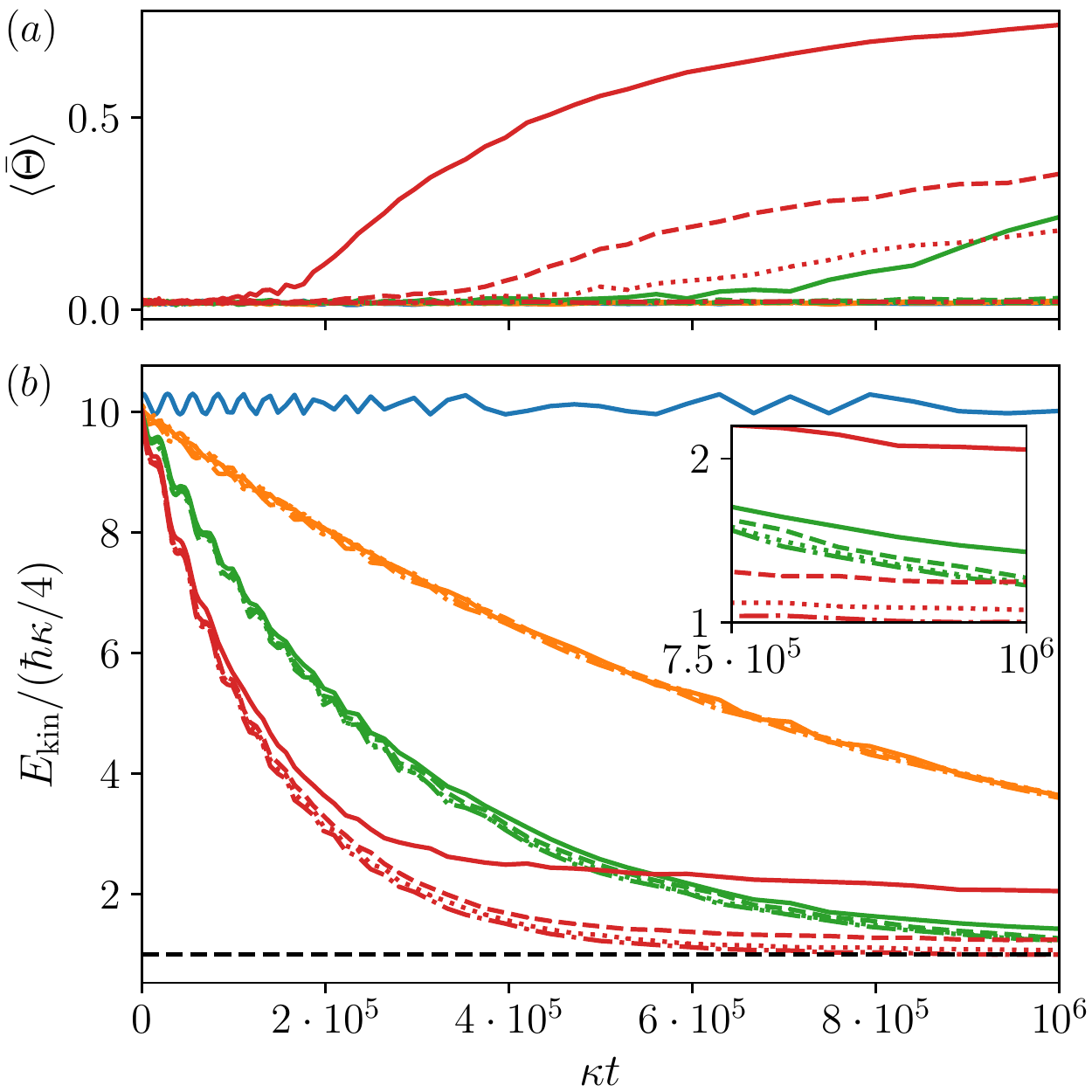}
\caption{Dynamics of (a) the total order parameter $\langle \bar \Theta \rangle$ and (b) the kinetic energy $E_\mathrm{kin}$ for $N=1000$ with an initial temperature $k_\mathrm{B} T_\mathrm{ini} = 10 \hbar \kappa / 2$ for different mode numbers (or bandwidths) $M=1$ (solid), $M=2$ (dashed), $M=10$ (dotted) and $M=50$ (dashdot). The depicted quantities are averages over $25$ trajectories. While for $\zeta_\mathrm{tot} = 0$ (blue) there is obviously no cooling, a non-zero pump strength leads to cooling. When pumping below the multi- and single-mode threshold $\zeta_\mathrm{tot} = 0.5$ (orange) there is no ordering and no difference in the kinetic energy curves for different mode numbers (or bandwidths). For $\zeta_\mathrm{tot} = 1.5$ (green), below the multi-mode threshold, but above the single-mode threshold we observe slower cooling for the $M=1$ and concurrent ordering. This difference becomes large for $\zeta_\mathrm{tot} = 3$ (red).}
\label{fig:cooling}
\end{figure}

A central application of the coupled atom field dynamics is motional cooling without spontaneous emission \cite{domokos2003mechanical}, which is known to have some important restrictions for larger ensembles \cite{hosseini2017cavity}.  Here we finally focus on the cooling capabilities in the extreme broadband case for different bandwidths and pump strengths. For this we simulate the dynamics of $N=1000$ atoms, starting with an initial temperature $k_\mathrm{B} T_\mathrm{ini} = 10 \hbar \kappa / 2$ which is ten times higher than the expected stationary temperature. Moreover we add a harmonic trap with the frequency $\omega_\mathrm{T} = 0.046 \omega_\mathrm{R}^{k_c}$ preventing the atomic cloud from expanding when they are not trapped by the self-consistent cavity potential. This is done by adding a restoring force $-m_a \omega_\mathrm{T}^2 x$ in Eq. \eqref{eq:semiclassical_momentum}.

\begin{figure}
\centering
\includegraphics[width=0.6\columnwidth]{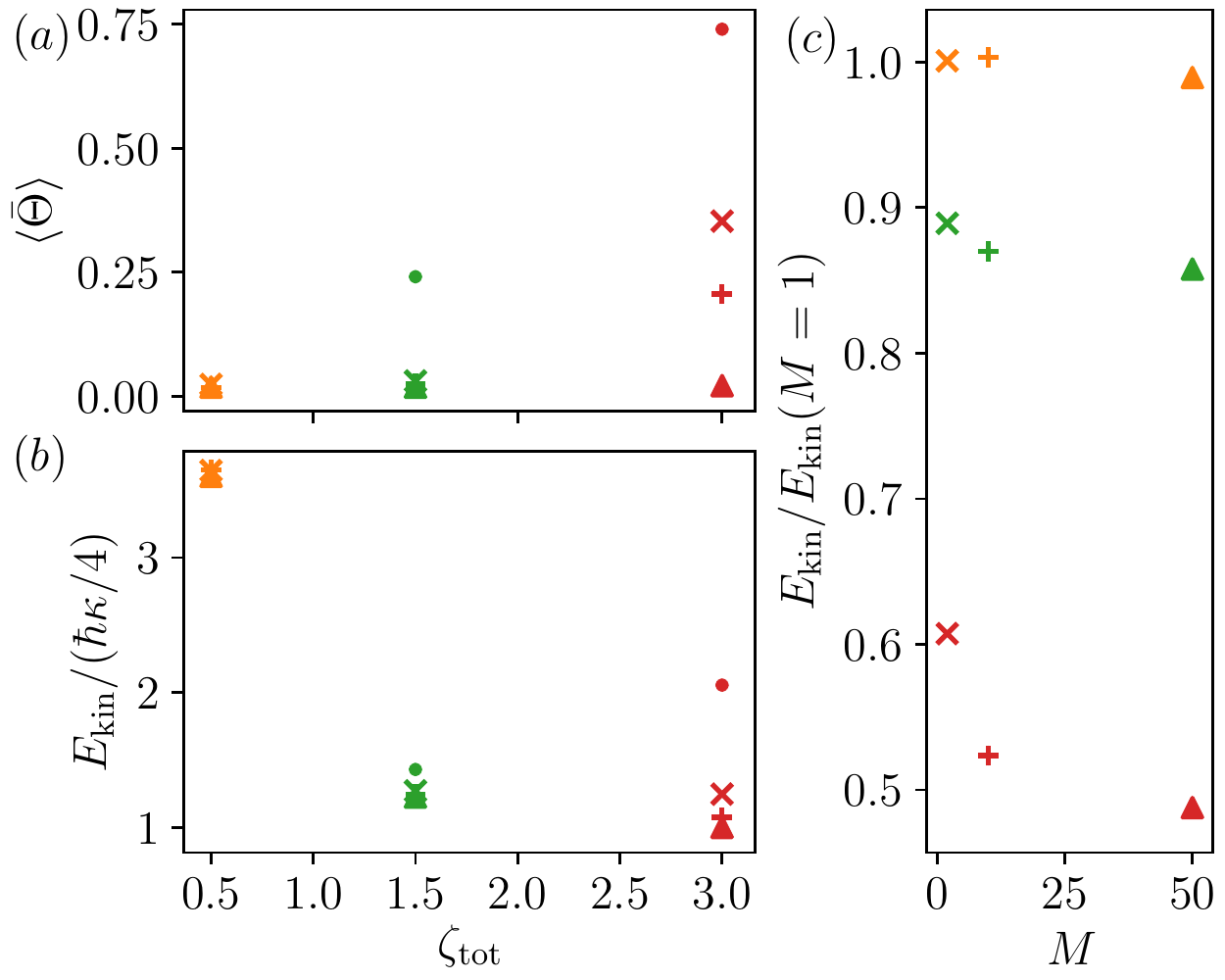}
\caption{The values of (a) the total order parameter $\langle \bar \Theta \rangle$, (b) kinetic energy $E_\mathrm{kin}$ and (c) the cooling enhancement compared to a single mode $E_\mathrm{kin} / E_\mathrm{kin}(M=1)$ at $\kappa t = 10^6$ in Fig.\ \ref{fig:cooling}. The symbols discriminate different mode numbers (or bandwidths): $M=1$ (disc), $M=2$ (cross), $M=10$ (plus) and $M=50$ (triangle). The colors depict the different pump intensities $\zeta_\mathrm{tot} = 0.5$ (orange), $1.5$ (green) and $3$ (red). Note that the stationary state has not been reached in all cases (see Fig.\ \ref{fig:cooling}).}
\label{fig:cooling_final}
\end{figure}

The resulting dynamics is depicted in Fig.\ \ref{fig:cooling} for different pump strengths and bandwidths. We ask how the cooling speed differs depending on the bandwidth for a certain $\zeta_\mathrm{tot}$. When pumping below the multi- and single-mode threshold $\zeta_\mathrm{tot} = 0.5$ there is no ordering and the kinetic energy curves for different bandwidths (or mode numbers) coincide (see the orange curve). Below the multi-mode threshold, but above the single-mode threshold (for $\zeta_\mathrm{tot} = 1.5 < 2$) there is ordering for $M=1$ only. This leads to slower cooling for $M=1$ \cite{niedenzu2011kinetic} compared to $M=2$ or $M=50$, as can be seen from the zoomed inset. For $\zeta_\mathrm{tot} = 3$ (red curve) this anti-correlation between ordering and fast cooling becomes even more evident. It is made explicit in Fig.\ \ref{fig:cooling_final}, where we plot the final values as a function of the bandwidth. The kinetic energy at $\kappa t = 10^6$ is decreased by a factor of about $0.49$ for the bandwidth $\Delta k_\mathrm{max} / k_c \approx 0.0213 = 2.13\%$ at $M=50$ compared to the single-mode case. Note that the kinetic energy follows a damped oscillation. When the cloud does not organize fast enough, it expands freely and bounces back from the fixed harmonic trap, leading to the oscillations in the beginning.

As already known from previous work \cite{niedenzu2011kinetic} the cooling speed increases with larger pump strength until the self-organization threshold is traversed. Then the ordering of the atoms slows down cooling and, for high pump strengths, also increases the stationary temperature. Using a train of short pulses with a large bandwidth shifts up the (multi-mode) self-organization threshold, avoiding self-ordering up to large pump powers. This allows for implementing faster cavity cooling.

\section{Conclusions}
\label{sec:conclusion}
In this work we theoretically studied the dynamics of a cold cloud of atoms in a multi-mode cavity which is illuminated by a train of phase-stabilized ultrashort pulses with a wide frequency spectrum. For red cavity and atomic detuning (high field seeking particles), the lowest energy state of the system is connected to a spatial atomic distribution, which optimally scatters the majority of the pulse frequency components into the cavity. In order to fulfill this requirement, in addition to a periodic Bragg grating with the period of the wavelength known from single-mode self-organization, the atoms need to be localized on a length scale smaller than the pulse length to adapt optimally to the mode structure.

We have shown that the threshold power required for self-ordering increases when the spatial pulse width is smaller than the atomic distribution as it is not possible for all the particles to commonly scatter in phase and ordering to all modes becomes increasingly difficult. Thus even at optimal single-mode order, less power is transmitted to the cavity and a fracturing of the atomic density into several lobes appears. Moreover, in the corresponding simulations we identify a hysteresis hinting that the phase transition switches from second to first order in the extreme multi-mode regime. 

Dynamically, the atoms do not attain this localization themselves. For high pump powers, they freeze in the wavelength Bragg grating, while for weak pump powers the atomic cloud expands out of the in-phase region of the modes. On the other side such a spatial dephasing of the modes generally has a positive effect on the cooling time and temperature. Using a large bandwidth avoids self-ordering and the concurrent increase of kinetic energy at too low pump power, while cavity cooling itself is hardly affected by the bandwidth.

Since the transmission of the pulse into the cavity depends on the atomic pattern, the cavity output intensity as a function of time can be used to gain information about the longitudinal form and location of the atomic cloud. This is similar to an atomic kaleidoscope \cite{salzburger2002enhanced} but operated in the longitudinal direction in the time domain. The formation of such an output pulse train with different periodicity and shape than the input pulse can be viewed as a spontaneous formation of a time crystal. While many of the discussed phenomena might survive in the zero temperature regime of quantum gas cavity QED, new phenomena could be expected from direct inter-particle interactions. 

\paragraph{Acknowledgments.} This work was performed in the framework of the European Training Network ColOpt, which is funded by the European Union (EU) Horizon 2020 programme under the Marie Sk\l odowska-Curie action, grant agreement 721465. We acknowledge support from the OeAD and the Croatian Ministry of Science and Education under the project WTZ Croatia 2018-19 (Wissenschaftlich Technische Zusammenarbeit/Scientific \& Technological Cooperation). In addition, the authors acknowledge support from the Croatian Science Foundation (Project Frequency-Comb-induced OptoMechanics - IP-2014-09-7342).

\section*{References}


\begin{thebibliography}{10}

\bibitem{domokos2003mechanical}
P.~Domokos and H.~Ritsch, ``Mechanical effects of light in optical
  resonators,'' {\em Journal of the Optical Society of America B}, vol.~20,
  no.~5, pp.~1098--1130, 2003.

\bibitem{Ritsch2013cold}
H.~Ritsch, P.~Domokos, F.~Brennecke, and T.~Esslinger, ``Cold atoms in
  cavity-generated dynamical optical potentials,'' {\em Reviews of Modern
  Physics}, vol.~85, no.~2, p.~553, 2013.

\bibitem{horak1997cavity}
P.~Horak, G.~Hechenblaikner, K.~M. Gheri, H.~Stecher, and H.~Ritsch,
  ``Cavity-induced atom cooling in the strong coupling regime,'' {\em Physical
  Review Letters}, vol.~79, no.~25, p.~4974, 1997.

\bibitem{domokos2002collective}
P.~Domokos and H.~Ritsch, ``Collective cooling and self-organization of atoms
  in a cavity,'' {\em Physical Review Letters}, vol.~89, no.~25, p.~253003,
  2002.

\bibitem{schleier2011optomechanical}
M.~H. Schleier-Smith, I.~D. Leroux, H.~Zhang, M.~A. Van~Camp, and
  V.~Vuleti{\'c}, ``Optomechanical cavity cooling of an atomic ensemble,'' {\em
  Physical review letters}, vol.~107, no.~14, p.~143005, 2011.

\bibitem{wolke2012cavity}
M.~Wolke, J.~Klinner, H.~Ke{\ss}ler, and A.~Hemmerich, ``Cavity cooling below
  the recoil limit,'' {\em Science}, vol.~337, no.~6090, pp.~75--78, 2012.

\bibitem{hosseini2017cavity}
M.~Hosseini, Y.~Duan, K.~M. Beck, Y.-T. Chen, and V.~Vuleti{\'c}, ``Cavity
  cooling of many atoms,'' {\em Physical review letters}, vol.~118, no.~18,
  p.~183601, 2017.

\bibitem{delic2019cavity}
U.~Deli{\'c}, M.~Reisenbauer, D.~Grass, N.~Kiesel, V.~Vuleti{\'c}, and
  M.~Aspelmeyer, ``Cavity cooling of a levitated nanosphere by coherent
  scattering,'' {\em Physical review letters}, vol.~122, no.~12, p.~123602,
  2019.

\bibitem{chan2003observation}
H.~W. Chan, A.~T. Black, and V.~Vuleti{\'c}, ``Observation of
  collective-emission-induced cooling of atoms in an optical cavity,'' {\em
  Physical review letters}, vol.~90, no.~6, p.~063003, 2003.

\bibitem{nagy2010dicke}
D.~Nagy, G.~K{\'o}nya, G.~Szirmai, and P.~Domokos, ``Dicke-model phase
  transition in the quantum motion of a bose-einstein condensate in an optical
  cavity,'' {\em Physical review letters}, vol.~104, no.~13, p.~130401, 2010.

\bibitem{baumann2010dicke}
K.~Baumann, C.~Guerlin, F.~Brennecke, and T.~Esslinger, ``Dicke quantum phase
  transition with a superfluid gas in an optical cavity,'' {\em Nature},
  vol.~464, no.~7293, p.~1301, 2010.

\bibitem{mekhov2012quantum}
I.~B. Mekhov and H.~Ritsch, ``Quantum optics with ultracold quantum gases:
  towards the full quantum regime of the light--matter interaction,'' {\em
  Journal of Physics B}, vol.~45, no.~10, p.~102001, 2012.

\bibitem{mivehvar2017superradiant}
F.~Mivehvar, H.~Ritsch, and F.~Piazza, ``Superradiant topological peierls
  insulator inside an optical cavity,'' {\em Physical review letters},
  vol.~118, no.~7, p.~073602, 2017.

\bibitem{mivehvar2019emergent}
F.~Mivehvar, H.~Ritsch, and F.~Piazza, ``Emergent quasicrystalline symmetry in
  light-induced quantum phase transitions,'' {\em Physical Review Letters},
  vol.~123, no.~21, p.~210604, 2019.

\bibitem{torggler2019quantum}
V.~Torggler, P.~Aumann, H.~Ritsch, and W.~Lechner, ``A quantum n-queens
  solver,'' {\em Quantum}, vol.~3, p.~149, 2019.

\bibitem{torggler2017quantum}
V.~Torggler, S.~Kr{\"a}mer, and H.~Ritsch, ``Quantum annealing with ultracold
  atoms in a multimode optical resonator,'' {\em Physical Review A}, vol.~95,
  no.~3, p.~032310, 2017.

\bibitem{landini2018formation}
M.~Landini, N.~Dogra, K.~Kr{\"o}ger, L.~Hruby, T.~Donner, and T.~Esslinger,
  ``Formation of a spin texture in a quantum gas coupled to a cavity,'' {\em
  Physical review letters}, vol.~120, no.~22, p.~223602, 2018.

\bibitem{Mivehvar2019cavity}
F.~Mivehvar, H.~Ritsch, and F.~Piazza, ``Cavity-quantum-electrodynamical
  toolbox for quantum magnetism,'' {\em Physical Review Letters}, vol.~122,
  no.~11, p.~113603, 2019.

\bibitem{kroeze2019dynamical}
R.~M. Kroeze, Y.~Guo, and B.~L. Lev, ``Dynamical spin-orbit coupling of a
  quantum gas,'' {\em arXiv preprint arXiv:1904.08388}, 2019.

\bibitem{kroeze2018spinor}
R.~M. Kroeze, Y.~Guo, V.~D. Vaidya, J.~Keeling, and B.~L. Lev, ``Spinor
  self-ordering of a quantum gas in a cavity,'' {\em Physical review letters},
  vol.~121, no.~16, p.~163601, 2018.

\bibitem{domokos2002dissipative}
P.~Domokos, T.~Salzburger, and H.~Ritsch, ``Dissipative motion of an atom with
  transverse coherent driving in a cavity with many degenerate modes,'' {\em
  Physical Review A}, vol.~66, no.~4, p.~043406, 2002.

\bibitem{nimmrichter2010master}
S.~Nimmrichter, K.~Hammerer, P.~Asenbaum, H.~Ritsch, and M.~Arndt, ``Master
  equation for the motion of a polarizable particle in a multimode cavity,''
  {\em New Journal of Physics}, vol.~12, no.~8, p.~083003, 2010.

\bibitem{holzmann2016tailored}
D.~Holzmann and H.~Ritsch, ``Tailored long range forces on polarizable
  particles by collective scattering of broadband radiation,'' {\em New Journal
  of Physics}, vol.~18, no.~10, p.~103041, 2016.

\bibitem{keller2018quenches}
T.~Keller, V.~Torggler, S.~B. J{\"a}ger, S.~Sch{\"u}tz, H.~Ritsch, and
  G.~Morigi, ``Quenches across the self-organization transition in multimode
  cavities,'' {\em New Journal of Physics}, vol.~20, no.~2, p.~025004, 2018.

\bibitem{niedenzu2011kinetic}
W.~Niedenzu, T.~Grie{\ss}er, and H.~Ritsch, ``Kinetic theory of cavity cooling
  and self-organisation of a cold gas,'' {\em Europhysics Letters}, vol.~96,
  no.~4, p.~43001, 2011.

\bibitem{schuetz2015thermodynamics}
S.~Sch{\"u}tz, S.~B. J{\"a}ger, and G.~Morigi, ``Thermodynamics and dynamics of
  atomic self-organization in an optical cavity,'' {\em Physical Review A},
  vol.~92, no.~6, p.~063808, 2015.

\bibitem{keller2017phases}
T.~Keller, S.~B. J{\"a}ger, and G.~Morigi, ``Phases of cold atoms interacting
  via photon-mediated long-range forces,'' {\em Journal of Statistical
  Mechanics: Theory and Experiment}, vol.~2017, no.~6, p.~064002, 2017.

\bibitem{torggler2014adaptive}
V.~Torggler and H.~Ritsch, ``Adaptive multifrequency light collection by
  self-ordered mobile scatterers in optical resonators,'' {\em Optica}, vol.~1,
  no.~5, pp.~336--342, 2014.

\bibitem{leonard2017supersolid}
J.~L{\'e}onard, A.~Morales, P.~Zupancic, T.~Esslinger, and T.~Donner,
  ``Supersolid formation in a quantum gas breaking a continuous translational
  symmetry,'' {\em Nature}, vol.~543, no.~7643, pp.~87--90, 2017.

\bibitem{gangl2000cold}
M.~Gangl and H.~Ritsch, ``Cold atoms in a high-{Q} ring cavity,'' {\em Physical
  Review A}, vol.~61, no.~4, p.~043405, 2000.

\bibitem{asboth2005self}
J.~Asb{\'o}th, P.~Domokos, H.~Ritsch, and A.~Vukics, ``Self-organization of
  atoms in a cavity field: Threshold, bistability, and scaling laws,'' {\em
  Physical Review A}, vol.~72, no.~5, p.~053417, 2005.

\bibitem{schuetz2014prethermalization}
S.~Sch{\"u}tz and G.~Morigi, ``Prethermalization of atoms due to
  photon-mediated long-range interactions,'' {\em Physical Review Letters},
  vol.~113, no.~20, p.~203002, 2014.

\bibitem{salzburger2002enhanced}
T.~Salzburger, P.~Domokos, and H.~Ritsch, ``Enhanced atom capturing in a high-q
  cavity by help of several transverse modes,'' {\em Optics Express}, vol.~10,
  no.~21, pp.~1204--1214, 2002.

\end{thebibliography}

\end{document}